\documentclass[conference]{IEEEtran}
\ifCLASSINFOpdf
\else
\fi
\usepackage{fancybox,ascmac}
\usepackage{amsmath}
\usepackage{latexsym}
\usepackage{multicol}
\usepackage[dvipdfmx]{graphicx}
\usepackage{mediabb}
\usepackage{amsmath, amssymb}
\usepackage{type1cm}
\usepackage{cases}
\usepackage{color}
\usepackage{array,booktabs}
\usepackage{slashbox}
\usepackage{here}
\usepackage{comment}

\allowdisplaybreaks

\newtheorem{definition}{Definition}
\newtheorem{theorem}{Theorem}
\newtheorem{lemma}{Lemma}
\newtheorem{remark}{Remark}

\begin{document}
%
\title{Variable-Length Intrinsic Randomness Allowing Positive Value of the Average Variational Distance}

\author{\IEEEauthorblockN{Jun Yoshizawa}
\IEEEauthorblockA{Waseda University \\
                    Email: junbadchel0313@suou.waseda.jp}
\and
\IEEEauthorblockN{Shota Saito}
\IEEEauthorblockA{Waseda University \\
                    Email: shota@aoni.waseda.jp}
\and
\IEEEauthorblockN{Toshiyasu Matsushima}
\IEEEauthorblockA{Waseda University \\
                    Email: toshimat@waseda.jp}}


%


\maketitle

\begin{abstract}
This paper considers the problem of variable-length intrinsic randomness. We propose the average variational distance as the performance criterion from the viewpoint of a dual relationship with the problem formulation of variable-length resolvability. Previous study has derived the general formula of the $\epsilon$-variable-length resolvability. We derive the general formula of the $\epsilon$-variable-length intrinsic randomness. Namely, we characterize the supremum of the mean length under the constraint that the value of the average variational distance is smaller than or equal to a constant $\epsilon$. Our result clarifies a dual relationship between the general formula of $\epsilon$-variable-length resolvability and that of $\epsilon$-variable-length intrinsic randomness. We also derive a lower bound of the quantity characterizing our general formula. 
\end{abstract}


%
\IEEEpeerreviewmaketitle

\section{Introduction}

The problem of random number generation is one of the important research topics in {\it Shannon theory}.
This problem is divided into 
\begin{itemize}
\item[(i)] the problem of resolvability (e.g. \cite{han1}, \cite{han3}, \cite{uyematsu1}, \cite{yagi}, \cite{yagi2}),
\item[(ii)] the problem of intrinsic randomness (e.g. \cite{han1}, \cite{han2}, \cite{uyematsu2}, \cite{vembu}).
\end{itemize}
For these problems, the {\it variational distance} is a major criterion used to measure the difference between the probability distribution generated by a mapping from a {\it coin distribution} \cite{han1} and a {\it target distribution} \cite{han1}.
Further, those problems are divided into 
\begin{itemize}
\item[(A)] the case of fixed-length (e.g. \cite{han1}, \cite{han3}, \cite{uyematsu1}, \cite{uyematsu2}),
\item[(B)] the case of variable-length (e.g. \cite{han1}, \cite{han2}, \cite{vembu}, \cite{yagi}, \cite{yagi2}).
\end{itemize} 

Investigating a duality between resolvability and intrinsic randomness is one of the important research topics.
For the problems of fixed-length resolvability ((i) \& (A)) and fixed-length intrinsic randomness ((ii) \& (A)),
a duality of those general formulae has been studied.
One way to capture the dual relationship of the general formulae is to see them from the viewpoint of the {\it smooth R\'enyi entropy} \cite{renner}.
For the problem of fixed-length resolvability, Uyematsu \cite{uyematsu1} has characterized the general formula by using the {\it smooth R\'enyi entropy of order zero} \cite{renner}. On the other hand, for the problem of fixed-length intrinsic randomness, Uyematsu and Kunimatsu \cite{uyematsu2} have characterized the general formula by using the {\it smooth R\'enyi entropy of order infinity} \cite{renner}.

For the problem of variable-length resolvability ((i) \& (B)), 
Yagi and Han \cite{yagi}, \cite{yagi2} have characterized the infimum of the mean length allowing positive value of the variational distance. 
However, the dual problem formulation to this problem has not been discussed yet.

This paper considers the problem of variable-length intrinsic randomness ((ii) \& (B))
and discusses the duality with the work by Yagi and Han \cite{yagi}, \cite{yagi2}.
From the viewpoint of a dual relationship with the problem formulation of variable-length resolvability, 
we propose the average variational distance.
This is the expectation of the variational distance between the probability distribution generated by a mapping and the uniform distribution for each length, where the expectation is taken with respect to the length. 
As the main result, we characterize the supremum of the mean length allowing positive value of the average variational distance.

We can see a duality between the general formula by Yagi and Han \cite{yagi}, \cite{yagi2} and our general formula from the viewpoint of the smooth R\'enyi entropy.
The general formula of Yagi and Han \cite{yagi}, \cite{yagi2} is related to the {\it smooth R\'enyi entropy of order $\alpha \in (0,1)$} \cite{renner} (cf. \cite{koga1}, \cite{koga2}, \cite{yagi2}). On the other hand, our general formula is related to the sub-probability distribution which achieves the infimum of the {\it smooth R\'enyi entropy of order $\alpha \in (1, \infty)$} \cite{renner} (cf. \cite{koga1}).

It is worth noticing that our problem formulation is different from the original formulation introduced by Vembu and Verd\'u \cite{vembu}. Vembu and Verd\'u \cite{vembu} and Han \cite{han1}, \cite{han2} have derived the general formula of the supremum of the mean length under the constraint that the value of the variational distance is equal to zero. 
Their variational distance measures the supremum of the difference between a {\it conditional probability distribution} generated by a mapping given each length and a uniform distribution. 
On the other hand, we consider one probability distribution on all lengths, because we consider the average variational distance. Therefore, our problem formulation is different from their problem formulation.

The organization of this paper is as follows. In Sec. II, we introduce the result of variable-length intrinsic randomness by Han \cite{han1}, \cite{han2}. In Sec. III, we state the result of $\epsilon$-variable-length resolvability by Yagi and Han \cite{yagi}, \cite{yagi2}. In Sec. I\hspace{-.1em}V, we describe $\epsilon$-variable-length intrinsic randomness. Specifically, Sec. I\hspace{-.1em}V-A introduces the problem formulation of our study. In Sec. I\hspace{-.1em}V-B, we describe the general formula of the $\epsilon$-variable-length intrinsic randomness. Further, we state a lower bound of the quantity characterizing the $\epsilon$-variable-length intrinsic randomness. In Sec. V, we prove our results. In Sec. V\hspace{-.1em}I, we discuss the dual relationship between the result by Yagi and Han \cite{yagi}, \cite{yagi2} and our result. In Sec. V\hspace{-.1em}II, we consider the second-order general formula. Finally, in Sec. V\hspace{-.1em}III, we summarize this paper.

\section{Variable-Length Intrinsic Randomness: Review}

Let $\mathcal{X}$ be a finite or countably infinite alphabet and $\mathcal{X}^n$ be the $n$-th Cartesian product of $\mathcal{X}$. Let $X^n$ be a random variable taking a value in $\mathcal{X}^n$ and $x^n$ be a realization of $X^n$. Let $\mathbf{X} = \{ X^n \}_{n=1}^\infty$ be a {\it general source} \cite{han1}. We denote by $\mathcal{P}(\mathcal{X}^n)$ a set of probability distribution $P_{X^n}$ on $\mathcal{X}^n$. We do not impose any assumptions such as stationarity or ergodicity. Let $\mathcal{U} = \{ 0, 1, \dots , K-1\}$ be a finite alphabet of size $K$, where $K$ is an integer greater than or equal to 2. For any nonnegative integer $m$, $U^{(m)}$ denotes a random variable distributed uniformly on $\mathcal{U}^{m}$ and $u^{(m)}$ denotes a realization of $U^{(m)}$, where $m$ is called the {\it length} of $U^{(m)}$. Let $\mathcal{U}^*$ be the set of all finite strings taken from $\mathcal{U}$, including the null string $\Lambda$ whose length is zero, i.e., $\mathcal{U}^* = \{ \Lambda, 0, 1, 00, \dots \}$. A mapping $\varphi_n$ is defined as $\varphi_n: \mathcal{X}^n \to \mathcal{U}^*$. Let $l(\varphi_n(x^n))$ be the length of $\varphi_n(x^n)$. Given $m$ and $\varphi_n$, the set $\mathcal{D}_m$ is defined as 
\begin{align*}
\mathcal{D}_m = \{ x^n \in \mathcal{X}^n \mid l(\varphi_n(x^n)) = m \}.
\end{align*}
Given $\varphi_n$, the set $\mathcal{J}(\varphi_n)$ is defined as 
\begin{align*}
\mathcal{J}(\varphi_n) = \{ m \in \mathbb{Z}_{\ge 0} \mid \mathbb{P}[ l(\varphi_n(X^n)) = m] >0 \}, 
\end{align*}
where $\mathbb{Z}_{\ge 0}$ is the set of nonnegative integers. We denote by\footnote{In this paper, logarithms are of base $K$.} $\iota_{P_X}(x) := \log \frac{1}{P_{X}(x)}$.
The {\it variational distance} between two probability distributions $P_X$ and $Q_X$ is defined as $d(P_X, Q_X) := \frac{1}{2} \sum_{x \in \mathcal{X}}|P_X(x) - Q_X(x)|$. A probability distribution $P_{X_m^n}$ is defined as 
\begin{align*}
P_{X_m^n}(x^n) = \frac{P_{X^n}(x^n)}{\mathbb{P}[X^n \in \mathcal{D}_m]} \ \ \ (x^n \in \mathcal{D}_m). 
\end{align*}

Previous studies such as \cite{han1}, \cite{han2}, and \cite{vembu} investigated the {\it problem of variable-length intrinsic randomness} defined as follows.

\begin{definition}[\cite{han1}, \cite{han2}]
{\rm A rate $R$ is said to be} {\it i$\mathchar`-$achievable} if there exists a mapping $\varphi_n:\mathcal{X}^n \to \mathcal{U}^*$ {\rm satisfying} 
\begin{align}
\limsup_{n\to \infty} \sup_{m \in \mathcal{J}(\varphi_n)} d(P_{\varphi_n(X_m^n)}, P_{U^{(m)}}) &= 0, \label{hantag} \\
\liminf_{n \to \infty} \frac{1}{n} \mathbb{E}_{P_{X^n}} [l(\varphi_n(X^n))] &\ge R, \notag
\end{align}
{\rm where $\mathbb{E}_{P_{X^n}}[\cdot]$ denotes the expectation with respect to the distribution $P_{X^n}$.}
\end{definition}

The {\it variable-length intrinsic randomness} \cite{han1}, \cite{han2} is defined as follows.
\begin{definition}[\cite{han1}, \cite{han2}]
\begin{align*}
S_{\rm i}(\mathbf{X}) := \sup \{ R \mid R \ {\rm is } \ {\rm i}\mathchar`-{\rm achievable}\}.
\end{align*}
\end{definition}

The following result was given by Han \cite{han1}, \cite{han2}.

\begin{theorem}[\cite{han1}, \cite{han2}]
For any general source $\mathbf{X}$,
\begin{align*}
S_{\rm i}(\mathbf{X}) = \liminf_{n \to \infty} \frac{1}{n} H(P_{X^n}), 
\end{align*}
where $H(P_{X^n})$ is the {\it entropy}.
\end{theorem}

\section{$\epsilon$-Variable-Length Resolvability: Review}

Let $L_n$ be a random variable taking a length $m$. Let $U^{(L_n)}$ be the {\it variable-length uniform random number} \cite{yagi}, \cite{yagi2}, where the probability distribution is defined as 
\begin{align*}
P_{U^{(L_n)}}(u^{(m)}, m) = \mathbb{P}[U^{(L_n)} \!=\! u^{(m)}, L_n = m] = \frac{\mathbb{P}[L_n = m]}{K^m}, 
\end{align*}
for all $u^{(m)} \in \mathcal{U}^m$. Therefore, $U^{(m)}$ is uniformly distributed over $\mathcal{U}^m$ given $L_n = m$. A mapping $\phi_n$ is defined as $\phi_n: \mathcal{U}^* \to \mathcal{X}^n$.

Previous studies such as \cite{yagi} and \cite{yagi2} investigated the {\it problem of $\epsilon$-variable-length resolvability} defined as follows.

\begin{definition}[\cite{yagi}, \cite{yagi2}]
{\rm Given $\epsilon \in [0,1)$, a rate $R$ is said to be }{\it r($\epsilon$)-achievable} {\rm if there exists a variable-length uniform random number} $U^{(L_n)}$ {\rm and a mapping} $\phi_n:\mathcal{U}^* \to \mathcal{X}^n$ {\rm satisfying} 
\begin{align*}
\limsup_{n\to \infty} d(P_{\phi_n(U^{(L_n)})}, P_{X^n}) &\le \epsilon, \\
\limsup_{n \to \infty} \frac{1}{n} \mathbb{E}_{P_{L_n}} [L_n] &\le R. 
\end{align*}
\end{definition}

The $\epsilon${\it -variable-length resolvability} \cite{yagi}, \cite{yagi2} is defined as follows.

\begin{definition}[\cite{yagi}, \cite{yagi2}]
\begin{align*}
S_{\rm r}(\epsilon | \mathbf{X}) := \inf \{ R \mid R \ {\rm is \ r}(\epsilon)\mathchar`-\rm{achievable}\}.
\end{align*}
\end{definition}

The following quantity was defined by Koga and Yamamoto \cite{koga2}.
\begin{definition}[\cite{koga2}]
Given $\epsilon \in [0,1)$, $G_{[\epsilon]}(\mathbf{X})$ is defined as
\begin{align*}
G_{[\epsilon]}(\mathbf{X}) = \lim_{\tau \downarrow 0} \limsup_{n \to \infty} \frac{1}{n} G_{[\epsilon + \tau]}(X^n),
\end{align*}
where $G_{[\epsilon + \tau]}(X^n)$ is defined as
\begin{align*}
G_{[\epsilon + \tau]}(X^n) = \inf_{A_n : \mathbb{P}[X^n \in A_n] \ge 1-\epsilon - \tau} \sum_{x^n \in A_n} &P_{X^n}(x^n) \\
&\cdot \log \frac{\mathbb{P}[X^n \in A_n]}{P_{X^n}(x^n)}.
\end{align*}
\end{definition}

The following result was given by Yagi and Han \cite{yagi}, \cite{yagi2} (cf. \cite{koga2}).

\begin{theorem}[\cite{yagi}, \cite{yagi2}]
For any general source $\mathbf{X}$,
\begin{align*}
S_{\rm r}(\epsilon | \mathbf{X}) = G_{[\epsilon]}(\mathbf{X}) \ \ \ (\epsilon \in [0, 1)).
\end{align*}
\end{theorem}

\begin{remark}
The study \cite{koga2} derived the general formula of {\it weak variable-length source coding} allowing {\it $\epsilon$-error probability}. The general formula is also characterized by $G_{[\epsilon]}(\mathbf{X})$.
\end{remark}

\section{$\epsilon$-Variable-Length Intrinsic Randomness}

\subsection{Problem Formulation}

In this study, let $\mathcal{X}$ be a finite alphabet. In the problem of variable-length intrinsic randomness, the probability distribution of $L_n$ is defined as $P_{L_n}(m) = \mathbb{P}[X^n \in \mathcal{D}_m]$. Therefore, the probability distribution of $U^{(L_n)}$ is defined as 
\begin{align*}
P_{U^{(L_n)}}(u^{(m)}, m) \!=\! \mathbb{P}[U^{(L_n)} \!=\! u^{(m)}, L_n \!=\! m] \!=\! \frac{\mathbb{P}[X^n \in \mathcal{D}_m]}{K^m}, 
\end{align*}
for all $u^{(m)} \in \mathcal{U}^m$. 
The performance criteria are the {\it average variational distance} and the mean length. The average variational distance between $P_{\varphi_n(X^n)}$ and $P_{U^{(L_n)}}$ is defined as 
\begin{align*}
&\bar{d}(P_{\varphi_n(X^n)}, P_{U^{(L_n)}}) 
\\
& \ \ \ = \sum_{m \in \mathcal{J}(\varphi_n)} P_{L_n}(m) \frac{1}{2}\sum_{u \in \mathcal{U}^m} |P_{\varphi_n(X_m^n)}(u) - P_{U^{(m)}}(u) | \\
& \ \ \ = \frac{1}{2}\sum_{m \in \mathcal{J}(\varphi_n)} \mathbb{P}[X^n \in \mathcal{D}_m] \sum_{u \in \mathcal{U}^m} \biggl|\frac{P_{\varphi_n(X^n)}(u)}{\mathbb{P}[X^n \in \mathcal{D}_m]} - \frac{1}{K^m} \biggr| \\
& \ \ \ = \frac{1}{2}\sum_{m \in \mathcal{J}(\varphi_n)} \sum_{u \in \mathcal{U}^m} \biggl|P_{\varphi_n(X^n)}(u) - \frac{\mathbb{P}[X^n \in \mathcal{D}_m]}{K^m} \biggr|.
\end{align*}

We define the {\it problem of $\epsilon$-variable-length intrinsic randomness}.

\begin{definition}
{\rm Given $\epsilon \in [0,1)$, a rate $R$ is said to be }{\it i($\epsilon)$-achievable} {\rm if there exists a mapping} $\varphi_n:\mathcal{X}^n \to \mathcal{U}^*$ {\rm satisfying} 
\begin{align}
\limsup_{n\to \infty} \bar{d}(P_{\varphi_n(X^n)}, P_{U^{(L_n)}}) &\le \epsilon, \label{hyoukakijun1} \\
\liminf_{n \to \infty} \frac{1}{n} \mathbb{E}_{P_{X^n}} [l(\varphi_n(X^n))] &\ge R. \label{hyoukakijun2}
\end{align}
\end{definition}

\begin{remark}
The variational distance (\ref{hantag}) in Definition 1 measures the supremum of the difference between a conditional probability distribution given each length generated by a mapping and a uniform distribution. 
On the other hand, the average variational distance (\ref{hyoukakijun1}) in Definition 6 measures the difference between a probability distribution generated by a mapping and a probability distribution of variable-length uniform random number $U^{(L_n)}$. Therefore, unlike Definition 1, we consider one probability distribution on all lengths in Definition 6.
\end{remark}

The $\epsilon${\it -variable-length intrinsic randomness} is defined as follows.
\begin{definition}
\begin{align*}
S_{\rm i}(\epsilon | \mathbf{X}) := \sup \{ R \mid R \ {\rm is } \ {\rm i}(\epsilon)\mathchar`-\rm{achievable}\}.
\end{align*}
\end{definition}

Our concern is to investigate $S_{\rm i}(\epsilon | \mathbf{X})$ for a general source $\mathbf{X}$.

\subsection{Main Results}

The following set plays an important role in producing our main results.

\begin{definition}
Given $\delta \in [0,1)$, $\mathcal{Q}_{\delta}(\mathcal{X}^n)$ is defined as the set of {\it sub-probability distribution} $Q_{X^n}$ satisfying the following conditions:
\begin{align*}
&Q_{X^n}(x^n) > 0, \ (\forall x^n \in \{ x^n \in \mathcal{X}^n \mid P_{X^n}(x^n) > 0 \}), \\
&Q_{X^n}(x^n) \le P_{X^n}(x^n), \ (\forall x^n \in \mathcal{X}^n), \\
&\sum_{x^n \in \mathcal{X}^n} Q_{X^n}(x^n) = 1-\delta.
\end{align*}
\end{definition}

Next, using $\mathcal{Q}_{\epsilon}(\mathcal{X}^n)$, we introduce a new quantity.

\begin{definition}
Given $\epsilon \in [0,1)$, $G^{[\epsilon]}(\mathbf{X})$ is defined as
\begin{align*}
G^{[\epsilon]}(\mathbf{X}) = \lim_{\tau \downarrow 0} \liminf_{n \to \infty} \frac{1}{n} G^{[\epsilon + \tau]}(X^n),
\end{align*}
where $G^{[\epsilon + \tau]}(X^n)$ is defined as
\begin{align*}
G^{[\epsilon + \tau]}(X^n) = \sup_{Q_{X^n} \in \mathcal{Q}_{\epsilon + \tau}(\mathcal{X}^n)} \mathbb{E}_{P_{X^n}}[\iota_{Q_{X^n}}(X^n)].
\end{align*}
\end{definition}

The following theorem is the main result in this paper.

\begin{theorem}
For any general source $\mathbf{X}$,
\begin{align*}
S_{\rm i}(\epsilon | \mathbf{X}) = G^{[\epsilon]}(\mathbf{X}) \ \ \ (\epsilon \in [0, 1)).
\end{align*} 
\end{theorem}

\begin{IEEEproof}
The proofs of the direct part and the converse part are in Section V-A and Section V-B, respectively.
\end{IEEEproof}

In Sec. V\hspace{-.1em}I, we discuss the dual relationship between the general formula in Theorem 2 by Yagi and Han \cite{yagi}, \cite{yagi2} and our general formula in Theorem 3.

\begin{remark}
Instead of (\ref{hyoukakijun1}), we consider the next condition:
\begin{align*}
\bar{d}(P_{\varphi_n(X^n)}, P_{U^{(L_n)}}) \le \epsilon \ \ \ (\forall n \ge n_0) 
\end{align*}
for some $n_0 \in \mathbb{N}$. We define $\tilde{S}_{\rm i}(\epsilon | \mathbf{X})$ as $\epsilon$-variable-length intrinsic randomness corresponding to this condition. Then, we have
\begin{align*}
\tilde{S}_{\rm i}(\epsilon | \mathbf{X}) = \liminf_{n \to \infty} \frac{1}{n} G^{[\epsilon]}(X^n). 
\end{align*}
\end{remark}

The following theorem characterizes the lower bound of $G^{[\epsilon]}(\mathbf{X})$.

\begin{theorem}
For any general source $\mathbf{X}$, 
\begin{align*}
\underline{H}_{\epsilon}(\mathbf{X}) \le G^{[\epsilon]}(\mathbf{X}) \ \ \ (\epsilon \in [0,1)),
\end{align*}
where the quantity of the lower bound is defined as 
\begin{align*}
\underline{H}_{\epsilon}(\mathbf{X}) = \sup \biggl\{ R \mid \limsup_{n \to \infty} \mathbb{P} \biggl[ \frac{1}{n} \iota_{P_{X^n}}(X^n) \le R \biggr] \le \epsilon \biggr\}.
\end{align*}
\end{theorem}

\begin{IEEEproof}
See Section V-C.
\end{IEEEproof}

\begin{remark}
The quantity $\underline{H}_{\epsilon}(\mathbf{X})$ characterizes $\epsilon$-{\it fixed-length intrinsic randomness} \cite{han1}.
\end{remark}

\section{Proofs of Main Results}

\subsection{Proof of the Direct Part of Theorem 3}

For any $\tau > 0$, $n \in \mathbb{N}$, and $\gamma > 0$, there exist a $\tilde{Q}_{X^n} \in \mathcal{Q}_{\epsilon+\tau}(\mathcal{X}^n)$ satisfying
\begin{align}
\mathbb{E}_{P_{X^n}} [\iota_{\tilde{Q}_{X^n}}(X^n)] &> \sup_{Q_{X^n} \in \mathcal{Q}_{\epsilon + \tau}(\mathcal{X}^n)} \mathbb{E}_{P_{X^n}}[\iota_{Q_{X^n}}(X^n)]  - \gamma \notag \\
&= G^{[\epsilon + \tau]}(X^n) - \gamma. \label{80}
\end{align}

\noindent
{\bf [Definitions of notation]}

\begin{itemize}
\item $R_j$ is defined as $R_j = 3\gamma j, \ (j = 0,1,2,\dots).$
\item $I_j$ is defined as $I_j = [R_j, R_{j+1}), \ (j = 0,1,2,\dots).$
\item $S_n^{(j)} \subset \mathcal{X}^n$ is defined as
\begin{align}
S_n^{(j)} \!=\! \biggl\{ x^n \!\in\! \mathcal{X}^n \mid \frac{1}{n} \iota_{\tilde{Q}_{X^n}}(x^n) \!\in\! I_j \biggr\}, \  (j = 0,1,2,\dots). \label{838}
\end{align}
\item $J$ is defined as $J = \{ 0,1,2, \dots \}.$
\item We partition $J$ into $J_1$ and $J_2$ defined as
\begin{align}
J_1 &= \{ j \ge 1 \mid \mathbb{P} [ X^n \in S_n^{(j)} ] \ge K^{-n\gamma R_j}\}, \label{940} \\ 
J_2 & = \{ 0 \} \cup \{ j \ge 1 \mid \mathbb{P} [ X^n \in S_n^{(j)} ] < K^{-n\gamma R_j}\}. \label{941}
\end{align}
\end{itemize}

For $x^n \in S_n^{(j)}$, it holds that
\begin{align*}
\tilde{Q}_{X^n}(x^n) \le K^{-nR_j} = K^{-n\gamma R_j} K^{-n(1-\gamma)R_j}. 
\end{align*}
Then, for $j \in J_1$ from (\ref{940}), it follows that
\begin{align}
\tilde{Q}_{X_n}(x^n) \le \mathbb{P}[X^n \in S_n^{(j)}]K^{-n(1-\gamma)R_j}. \label{944}
\end{align}

\noindent
{\bf [Construction of the mapping]}

We use the following Lemma 1. The proof of this lemma is similar to that of Lemma 2.2 in \cite{han2}.

\begin{lemma}
Let $R>0$, $a>0$ be any constants, $\gamma > 0$ be an arbitrarily small constant, $A_n \subset \mathcal{X}^n$ be an arbitrarily set, and $c \ge \mathbb{P}[X^n \in A_n]$ be an arbitrarily constant. Suppose that the probability distribution $P_{X^n} \in \mathcal{P}(\mathcal{X}^n)$ satisfies the condition
\begin{align}
P_{X^n}(x^n) \le cK^{-n(a+\gamma)R} \ \ \ (\forall x^n \in A_n). \label{2.1}
\end{align} 
Then, there exists a mapping $\varphi_n:A_n \to \mathcal{U}^{\lfloor naR \rfloor}$ such that 
\begin{align}
&\frac{1}{2} \sum_{u \in \mathcal{U}^{\lfloor naR \rfloor}} \biggl| P_{\varphi_n(X^n)}(u) - \frac{c}{K^{\lfloor naR \rfloor}} \biggr| \notag \\
& \ \ \ \le cK^{-n\gamma R} + \frac{1}{2} (c- \mathbb{P}[X^n \in A_n]). \label{hodai1}
\end{align}
\end{lemma}

\begin{IEEEproof}
See Appendix A.
\end{IEEEproof}

\begin{remark}
It is easy to check from the way of the proof of Lemma 1 that Lemma 1 holds even if we replace the probability distribution $P_{X^n}$ and $\mathbb{P}[X^n \in A_n]$ defined by $P_{X^n}$ by sub-probability distributions.
\end{remark}

We use Lemma 1 with $R = R_j$, $a = 1-2\gamma$, $A_n = S_n^{(j)}$, $c = \mathbb{P}[X^n \in S_n^{(j)}] (\ge \tilde{\mathbb{Q}}[X^n \in S_n^{(j)}])$. From (\ref{944}), there exists a mapping $\varphi_n^{(j)}:S_n^{(j)} \to \mathcal{U}^{\lfloor n(1 - 2\gamma) R_j \rfloor}$ such that 
\begin{align}
&\frac{1}{2} \sum_{u \in \mathcal{U}^{\lfloor n(1-2\gamma)R_j \rfloor}} \biggl| \tilde{Q}_{\varphi_n^{(j)}(X^n)} (u) - \frac{\mathbb{P}[X^n \in S_n^{(j)}]}{K^{\lfloor n(1-2\gamma)R_j \rfloor}} \biggr| \notag \\
&\!\le\! \mathbb{P}[X^n \in S_n^{(j)}] K^{-n \gamma R_j} \!+\! \frac{1}{2} (\mathbb{P}[X^n \in S_n^{(j)}] \!-\! \mathbb{\tilde{Q}}[X^n \in S_n^{(j)}]), \label{945}
\end{align} 
where $\mathbb{\tilde{Q}}[X^n \in S_n^{(j)}]) = \sum_{x^n \in S_n^{(j)}} \tilde{Q}_{X^n}(x^n)$.

Next, we construct the mapping $\varphi_n: \mathcal{X}^n \to \mathcal{U}^*$ by
\begin{align}
\varphi_n(x^n) := 
\begin{cases}
\varphi_n^{(j)} (x^n), \ \ \ (x^n \in S_n^{(j)}, j \in J_1), \\
\Lambda, \ \ \ (\rm{otherwise}). \label{946}
\end{cases}
\end{align}
Therefore, it holds that
\begin{align}
\mathcal{J}(\varphi_n) = \{ 0 \} \cup \{ \lfloor n(1-2\gamma)R_j \rfloor \mid j \in J_1 \}. \label{947}
\end{align}

\noindent
{\bf [Evaluation of the average variational distance]}

From $\tilde{Q}_{X^n} \in \mathcal{Q}_{\epsilon+\tau}(\mathcal{X}^n)$, there exists a $\{ \epsilon^{(j)} \}_{j=1}^{|J|}$, $\{ \tau^{(j)} \}_{j=1}^{|J|}$ such that
\begin{align}
\epsilon^{(1)} + \epsilon^{(2)} + \dots + \epsilon^{(|J|)} &= \epsilon, \label{original0} \\
\tau^{(1)} + \tau^{(2)} + \dots + \tau^{(|J|)} &= \tau, \label{original01} \\
\mathbb{P}[X^n \in S_n^{(j)}] - \mathbb{\tilde{Q}}[X^n \in S_n^{(j)}] &= \epsilon^{(j)} + \tau^{(j)}. \label{original1}
\end{align}
For $j \in J_1$, we have
\begin{align}
&\frac{1}{2} \sum_{u \in \mathcal{U}^{\lfloor n(1-2\gamma)R_j \rfloor}} \biggl| \tilde{Q}_{\varphi_n(X^n)} (u) - \frac{\mathbb{P}[X^n \in S_n^{(j)}]}{K^{\lfloor n(1-2\gamma)R_j \rfloor}} \biggr| \notag \\
& \ \overset{(\rm{a})}{\le} \mathbb{P}[X^n \!\in\! S_n^{(j)}] K^{-n \gamma R_j} \!+\! \frac{1}{2} (\mathbb{P}[X^n \!\in\! S_n^{(j)}] \!-\! \mathbb{\tilde{Q}}[X^n \!\in\! S_n^{(j)}]) \notag \\
& \ \overset{(\rm{b})}{=} \mathbb{P}[X^n \in S_n^{(j)}] K^{-n \gamma R_j} + \frac{\epsilon^{(j)} + \tau^{(j)}}{2}, \label{1055}
\end{align}
where (a) follows from (\ref{945}) and (\ref{946}), (b) follows from (\ref{original1}). Then, it holds that
\begin{align}
&\frac{1}{2} \sum_{u \in \mathcal{U}^{\lfloor n(1-2\gamma)R_j \rfloor}} \biggl| P_{\varphi_n(X^n)} (u) - \frac{\mathbb{P}[X^n \in S_n^{(j)}]}{K^{\lfloor n(1-2\gamma)R_j \rfloor}} \biggr| \notag \\
& \ \le \frac{1}{2} \sum_{u \in \mathcal{U}^{\lfloor n(1-2\gamma)R_j \rfloor}} | P_{\varphi_n(X^n)} (u) - \tilde{Q}_{\varphi_n(X^n)}(u) | \notag \\
& \ \ \ + \frac{1}{2} \sum_{u \in \mathcal{U}^{\lfloor n(1-2\gamma)R_j \rfloor}} \biggl| \tilde{Q}_{\varphi_n(X^n)}(u) - \frac{\mathbb{P}[X^n \in S_n^{(j)}]}{K^{\lfloor n(1-2\gamma)R_j \rfloor}} \biggr| \notag \\
& \ \overset{(\rm{c})}{\le} \frac{1}{2} \sum_{u \in \mathcal{U}^{\lfloor n(1-2\gamma)R_j \rfloor}} | P_{\varphi_n(X^n)} (u) - \tilde{Q}_{\varphi_n(X^n)}(u)| \notag \\
& \ \ \ + \mathbb{P}[X^n \in S_n^{(j)}] K^{-n \gamma R_j} + \frac{\epsilon^{(j)} + \tau^{(j)}}{2},  \label{1058}
\end{align}
where (c) follows from (\ref{1055}). On the other hand, for $j \in J_2$, it follows that
\begin{align}
&\frac{1}{2} \sum_{u \in \mathcal{U}^0} \biggl| P_{\varphi_n(X^n)} (u) - \frac{\mathbb{P}[X^n \in \mathcal{D}_0]}{K^0} \biggr| \notag \\
& \ \ \ = \frac{1}{2} |P_{\varphi_n(X^n)} (\Lambda) - \mathbb{P}[X^n \in \mathcal{D}_0]| \notag \\
& \ \ \ = \frac{1}{2} | \mathbb{P}[X^n \in \mathcal{D}_0] - \mathbb{P}[X^n \in \mathcal{D}_0] | = 0. \label{1053}
\end{align}

Next, for sufficient large number $n \in \mathbb{N}$, we prove that the length $\lfloor n(1-2\gamma)R_{j} \rfloor$ differs for each $j \in J_1$. For $n \ge \frac{1}{(1-2\gamma)3\gamma}$, we have
\begin{align}
n(1-2\gamma)3\gamma(j+1) -1 \ge  n(1-2\gamma)3\gamma j . \label{t3}
\end{align}
From the definition of the {\it floor function}, it holds that
\begin{align}
\lfloor n(1-2\gamma)R_{j+1} \rfloor &> n(1-2\gamma)3\gamma(j+1) -1, \label{t4} \\
\lfloor n(1-2\gamma)R_{j} \rfloor &\le n(1-2\gamma)3\gamma j. \label{t5}
\end{align}
By substituting (\ref{t4}) and (\ref{t5}) for (\ref{t3}), we have
\begin{align*}
\lfloor n(1-2\gamma)R_{j+1} \rfloor > \lfloor n(1-2\gamma)R_{j} \rfloor.
\end{align*}
Therefore, we obtain the following fact: ($\spadesuit$) {\it For $n \ge \frac{1}{(1-2\gamma)3\gamma}$, the length $\lfloor n(1-2\gamma)R_{j} \rfloor$ differs for each $j \in J_1$.}

For $n \ge \frac{1}{(1-2\gamma)3\gamma}$, the combination of (\ref{946}), (\ref{947}), (\ref{1058}), (\ref{1053}), and ($\spadesuit$) yields
\begin{align}
&\bar{d}(P_{\varphi_n(X^n)}, P_{U^{(L_n)}}) \notag \\
& \ \ \ = \frac{1}{2} \sum_{m \in \mathcal{J}(\varphi_n)} \sum_{u \in \mathcal{U}^m} \biggl| P_{\varphi_n(X^n)} (u) - \frac{\mathbb{P}[X^n \in \mathcal{D}_m]}{K^m} \biggr| \notag \\
& \ \ \ = \frac{1}{2} \sum_{m \in \mathcal{J}(\varphi_n)} \sum_{u \in \mathcal{U}^m} \biggl| P_{\varphi_n(X^n)} (u) - \frac{\mathbb{P}[l(\varphi_n(X^n)) = m]}{K^m} \biggr| \notag \\
& \ \ \ = \frac{1}{2} \sum_{j \in J_1} \sum_{u \in \mathcal{U}^{\lfloor n(1-2\gamma)R_j \rfloor}} \biggl| P_{\varphi_n(X^n)} (u) - \frac{\mathbb{P}[X^n \in S_n^{(j)}]}{K^{\lfloor n(1-2\gamma)R_j \rfloor}} \biggr| \notag \\
& \ \ \ \le \frac{1}{2} \sum_{j \in J_1} \sum_{u \in \mathcal{U}^{\lfloor n(1-2\gamma)R_j \rfloor}} | P_{\varphi_n(X^n)} (u) - \tilde{Q}_{\varphi_n(X^n)}(u) | \notag \\
& \ \ \ \ \ + \sum_{j \in J_1} \mathbb{P}[X^n \in S_n^{(j)}] K^{-n \gamma R_j} + \sum_{j \in J_1} \frac{\epsilon^{(j)} + \tau^{(j)}}{2}. \label{original2}
\end{align}
First, we evaluate the first term on the right-hand side of (\ref{original2}).
\begin{align}
&\frac{1}{2} \sum_{j \in J_1} \sum_{u \in \mathcal{U}^{\lfloor n(1-2\gamma)R_j \rfloor}} | P_{\varphi_n(X^n)} (u) - \tilde{Q}_{\varphi_n(X^n)}(u) | \notag \\
& \ = \frac{1}{2} \sum_{j \in J_1} \sum_{u \in \mathcal{U}^{\lfloor n(1-2\gamma)R_j \rfloor}} \biggl| \sum_{x^n : \varphi_n(x^n) = u}(P_{X^n} (x^n) - \tilde{Q}_{X^n}(x^n)) \biggr| \notag \\
& \ \le \frac{1}{2} \sum_{j \in J_1} \sum_{u \in \mathcal{U}^{\lfloor n(1-2\gamma)R_j \rfloor}} \sum_{x^n : \varphi_n(x^n) = u}|P_{X^n} (x^n) - \tilde{Q}_{X^n}(x^n)| \notag \\
& \ \overset{(\rm{d})}{\le} \frac{1}{2} \sum_{x^n \in \mathcal{X}^n} |P_{X^n}(x^n) - \tilde{Q}_{X^n}(x^n)| \overset{(\rm{e})}{=} \frac{\epsilon + \tau}{2}, \label{1kou}
\end{align}
where (d) follows from (\ref{946}) and ($\spadesuit$), (e) follows from $\tilde{Q}_{X^n} \in \mathcal{Q}_{\epsilon + \tau}(\mathcal{X}^n)$. Next, we evaluate the second term on the right-hand side of (\ref{original2}).
\begin{align}
\sum_{j \in J_1} \mathbb{P}[X^n \in S_n^{(j)}] K^{-n \gamma R_j} &\le \sum_{j \in J \setminus \{ 0 \}} K^{-3n \gamma^2 j} \notag \\
&= \frac{K^{-3n\gamma^2}}{1 - K^{-3n\gamma^2}}. \label{2kou}
\end{align}
Finally, we evaluate the third term on the right-hand side of (\ref{original2}).
\begin{align}
\sum_{j \in J_1} \frac{\epsilon^{(j)} + \tau^{(j)}}{2} &\le \sum_{j \in J} \frac{\epsilon^{(j)} + \tau^{(j)}}{2} \overset{(\rm{f})}{=} \frac{\epsilon + \tau}{2}, \label{3kou}
\end{align}
where (f) follows from (\ref{original0}) and (\ref{original01}). By substituting (\ref{1kou}), (\ref{2kou}), and (\ref{3kou}) for (\ref{original2}), we have, for $n \ge \frac{1}{(1-2\gamma)3\gamma}$,
\begin{align*}
\bar{d}(P_{\varphi_n(X^n)}, P_{U^{(L_n)}}) &\le \epsilon + \tau + \frac{K^{-3n\gamma^2}}{1 - K^{-3n\gamma^2}}. 
\end{align*}
By letting $\tau \downarrow 0$, we have
\begin{align}
\limsup_{n \to \infty} \bar{d}(P_{\varphi_n(X^n)}, P_{U^{(L_n)}}) \le \epsilon. \label{1174}
\end{align}

\noindent
{\bf [Evaluation of the mean length]}

For $n \ge \frac{1}{(1-2\gamma)3\gamma}$, it follows that
\begin{align}
&\mathbb{E}_{P_{X^n}}[l(\varphi_n(X^n))] \notag \\
& \ = \sum_{m \in \mathcal{J}(\varphi_n)} m \mathbb{P} [l( \varphi_n(X^n)) = m] \notag \\
& \ \overset{(\rm{g})}{=} \sum_{j \in J_1} \lfloor n(1-2\gamma)R_{j} \rfloor \mathbb{P} [ X^n \in S_n^{(j)} ] \notag \\
& \ \ge \sum_{j \in J_1} (n(1-2\gamma)R_{j} -1) \mathbb{P} [ X^n \in S_n^{(j)} ] \notag \\
& \ \ge \sum_{j \in J_1} n(1-2\gamma)R_{j} \mathbb{P} [ X^n \in S_n^{(j)} ] -\sum_{j \in J} \mathbb{P} [ X^n \in S_n^{(j)} ] \notag \\
& \ = \sum_{j \in J_1} n(1-2\gamma)(R_{j+1}-3\gamma) \mathbb{P} [ X^n \in S_n^{(j)} ] - 1\notag \\
& \ \ge n(1-2\gamma)\sum_{j \in J_1}R_{j+1} \mathbb{P} [ X^n \in S_n^{(j)} ] - 3n\gamma(1 - 2\gamma) -1\notag \\
& \ = n(1-2\gamma)\sum_{j \in J}R_{j+1} \mathbb{P} [ X^n \in S_n^{(j)} ] \notag \\
& \ \ \ -n(1-2\gamma)\sum_{j \in J_2}R_{j+1} \mathbb{P} [ X^n \in S_n^{(j)} ] -3n\gamma(1 - 2\gamma) -1 \notag \\
& \ = n(1-2\gamma)\sum_{j \in J}R_{j+1} \mathbb{P} [ X^n \in S_n^{(j)} ] \notag \\
& \ \ \ -n(1-2\gamma)\sum_{j \in J_2 \setminus \{ 0\}}R_{j+1} \mathbb{P} [ X^n \in S_n^{(j)} ] \notag \\ & \ \ \ -n(1-2\gamma)R_1 \mathbb{P} [ X^n \in S_n^{(0)} ] - 3n\gamma(1 - 2\gamma) -1 \notag \\
& \ \ge n(1-2\gamma)\sum_{j \in J}R_{j+1} \mathbb{P} [ X^n \in S_n^{(j)} ] \notag \\
& \ \ \ \!-\!n(1-2\gamma)\sum_{j \in J_2 \setminus \{ 0\}}R_{j+1} \mathbb{P} [ X^n \!\in\! S_n^{(j)} ] \!-\!6n\gamma(1\!-\!2\gamma) \!-\!1, \label{1286}
\end{align}
where (g) follows from (\ref{946}). We evaluate the second term on the right-hand side of (\ref{1286}).
\begin{align}
&\sum_{j \in J_2 \setminus \{ 0\}}R_{j+1} \mathbb{P} [ X^n \in S_n^{(j)} ] 
\overset{(\rm{h})}{<} 3\gamma \sum_{j =1}^\infty (j+1) K^{-3n\gamma^2 j} \notag \\
& \ = \frac{3\gamma K^{-3n\gamma^2}}{1 \!-\! K^{-3n\gamma^2}} + \frac{3\gamma K^{-3n\gamma^2}}{(1 \!-\! K^{-3n\gamma^2})^2} \le \frac{6\gamma K^{-3n\gamma^2}}{(1 \!-\! K^{-3n\gamma^2})^2}, \label{oregairu}
\end{align}
where (h) follows from (\ref{941}). By substituting (\ref{oregairu}) for (\ref{1286}), it holds that
\begin{align}
&\frac{1}{n} \mathbb{E}_{P_{X^n}}[l(\varphi_n(X^n))] \ge (1-2\gamma)\sum_{j \in J}R_{j+1} \mathbb{P} [ X^n \in S_n^{(j)} ] \notag \\
& \ \ \ \ \ \ \ \ \ \ \ \ \  - \frac{6\gamma (1-2\gamma)K^{-3n\gamma^2}}{(1 - K^{-3n\gamma^2})^2} -6\gamma(1-2\gamma) -\frac{1}{n}. \label{1396}
\end{align}
We evaluate the first term on the right-hand side of (\ref{1396}).
\begin{align}
&(1-2\gamma)\sum_{j \in J}R_{j+1} \mathbb{P} [ X^n \in S_n^{(j)} ] \notag \\
& \ \overset{(\rm{i})}{>} (1-2\gamma)\sum_{j \in J} \sum_{x^n \in S_n^{(j)}} P_{X^n}(x^n) \frac{1}{n} \iota_{\tilde{Q}_{X^n}} (x^n) \notag \\
& \ = (1-2\gamma) \frac{1}{n} \mathbb{E}_{P_{X^n}} [\iota_{\tilde{Q}_{X^n}}(X^n)], \label{oregairu2}
\end{align}
where (i) follows from (\ref{838}). The combination of (\ref{oregairu2}) and (\ref{1396}) yields
\begin{align*}
&\frac{1}{n} \mathbb{E}_{P_{X^n}}[l(\varphi_n(X^n))] \\
& \ \ge (1-2\gamma) \frac{1}{n} \mathbb{E}_{P_{X^n}} [\iota_{\tilde{Q}_{X^n}}(X^n)] - \frac{6\gamma (1-2\gamma)K^{-3n\gamma^2}}{(1 - K^{-3n\gamma^2})^2} \\ & \ \ \ -6\gamma(1-2\gamma) -\frac{1}{n} \\
& \ \overset{(\rm{j})}{\ge} (1-2\gamma) \frac{1}{n} G^{[\epsilon + \tau]}(X^n) - \frac{(1-2\gamma)\gamma}{n}  \\ & \ \ \ - \frac{6\gamma (1-2\gamma)K^{-3n\gamma^2}}{(1 - K^{-3n\gamma^2})^2} -6\gamma(1-2\gamma) -\frac{1}{n},
\end{align*}
where (j) follows from (\ref{80}). Therefore, it follows that
\begin{align*}
&\liminf_{n \to \infty} \frac{1}{n} \mathbb{E}_{P_{X^n}}[l(\varphi_n(X^n))] \\
& \ \ge (1-2\gamma) \liminf_{n \to \infty} \frac{1}{n} G^{[\epsilon + \tau]}(X^n) -6\gamma(1-2\gamma).
\end{align*}
Since $\gamma > 0$ and $\tau > 0$ are arbitrary, this indicates that
\begin{align}
\liminf_{n \to \infty} \frac{1}{n} \mathbb{E}_{P_{X^n}}[l(\varphi_n(X^n))] &\ge G^{[\epsilon]}(\mathbf{X}). \label{13104}
\end{align}
From (\ref{1174}) and (\ref{13104}), $R$ satisfying $R < G^{[\epsilon]}(\mathbf{X})$ is i($\epsilon)$-achievable. Hence, we have $S_{\rm i}(\epsilon | \mathbf{X}) \ge G^{[\epsilon]}(\mathbf{X})$.

\subsection{Proof of the Converse Part of Theorem 3}

Suppose that $R$ is i$(\epsilon)\mathchar`-\rm{achievable}$, i.e., suppose that there exists a mapping $\varphi_n:\mathcal{X}^n \to \mathcal{U}^*$ satisfying (\ref{hyoukakijun1}) and (\ref{hyoukakijun2}). From (\ref{hyoukakijun1}), for $\tau > 0$, there exists an $n_0 \in \mathbb{N}$ such that 
\begin{align*}
\frac{1}{2} \sum_{m \in \mathcal{J}(\varphi_n)} \sum_{u \in \mathcal{U}^m} \biggl|P_{\varphi_n(X^n)}(u) - \frac{\mathbb{P}[X^n \in \mathcal{D}_m]}{K^m} \biggr| \le \epsilon + \tau
\end{align*}
for all $n \ge n_0$. There exists a sequence $\{ \epsilon^{(i)}\}_{i=1}^{|\mathcal{J}(\varphi_n)|}$, $\{ \tau^{(i)}\}_{i=1}^{|\mathcal{J}(\varphi_n)|}$ such that 
\begin{align}
\epsilon^{(1)} + \epsilon^{(2)} + \dots + \epsilon^{(|\mathcal{J}(\varphi_n)|)} &\le \epsilon,  \label{yui1} \\
\tau^{(1)} + \tau^{(2)} + \dots + \tau^{(|\mathcal{J}(\varphi_n)|)} &\le \tau, \label{yui2}
\end{align}
for $m \in \mathcal{J}(\varphi_n)$,
\begin{align*}
\frac{1}{2} \sum_{u \in \mathcal{U}^m} \biggl|P_{\varphi_n(X^n)}(u) - \frac{\mathbb{P}[X^n \in \mathcal{D}_m]}{K^m} \biggr| = \epsilon^{(m)} + \tau^{(m)}.
\end{align*}
Next, the set $A_n^{(m)} \subset \mathcal{X}^n$ for $m \in \mathcal{J}(\varphi_n)$ is defined as follows:
\begin{align*}
A_n^{(m)} \!=\! \biggl\{ x^n \!\in\! \mathcal{X}^n \mid P_{X^n}(x^n) \!\ge\! \frac{\mathbb{P}[X^n \!\in\! \mathcal{D}_m]}{K^m} , \varphi_n(x^n) \!\in\! \mathcal{U}^m \biggr\}.
\end{align*}
Moreover, we define the set $V_m \supset \varphi_n(A_n^{(m)})$ by
\begin{align*}
V_m = \biggl\{ u \in \mathcal{U}^m \mid P_{\varphi_n(X^n)}(u) \ge \frac{\mathbb{P}[X^n \in \mathcal{D}_m]}{K^m} \biggr\}.
\end{align*}

For $n \ge n_0$ and $m \in \mathcal{J}(\varphi_n)$, it holds that
\begin{align*}
&\epsilon^{(m)}+\tau^{(m)} = \frac{1}{2} \sum_{u \in \mathcal{U}^m} \biggl|P_{\varphi_n(X^n)}(u) - \frac{\mathbb{P}[X^n \in \mathcal{D}_m]}{K^m} \biggr| \\
&= \frac{1}{2} \sum_{u \in \mathcal{U}^m} \biggl|\sum_{x^n: \varphi_n(x^n) = u} P_{X^n}(x^n) - \frac{\mathbb{P}[X^n \in \mathcal{D}_m]}{K^m} \biggr| \\
&= \frac{1}{2} \sum_{u \in V_m} \biggl(\sum_{x^n: \varphi_n(x^n) = u} P_{X^n}(x^n) - \frac{\mathbb{P}[X^n \in \mathcal{D}_m]}{K^m} \biggr) \notag \\
& \ \ \ + \frac{1}{2} \sum_{u \in \mathcal{U}^m \setminus V_m} \biggl(\frac{\mathbb{P}[X^n \in \mathcal{D}_m]}{K^m} - \sum_{x^n: \varphi_n(x^n) = u}P_{X^n}(x^n) \biggr) \\
&= \frac{1}{2} \sum_{u \in V_m} \biggl(\sum_{x^n: \varphi_n(x^n) = u} P_{X^n}(x^n) - \frac{\mathbb{P}[X^n \in \mathcal{D}_m]}{K^m} \biggr) \notag \\
& \ \ \ + \frac{1}{2} \biggl(\mathbb{P}[X^n \in \mathcal{D}_m] - \sum_{u \in V_m} \frac{\mathbb{P}[X^n \in \mathcal{D}_m]}{K^m} \biggr) \notag \\
& \ \ \ - \frac{1}{2} \biggl(\mathbb{P}[X^n \in \mathcal{D}_m] - \sum_{u \in V_m} \sum_{x^n: \varphi_n(x^n) = u}P_{X^n}(x^n) \biggr) \\
&= \sum_{u \in V_m} \biggl(\sum_{x^n: \varphi_n(x^n) = u} P_{X^n}(x^n) - \frac{\mathbb{P}[X^n \in \mathcal{D}_m]}{K^m} \biggr) \\
&\overset{(\rm{k})}{\ge} \sum_{u \in \varphi_n(A_n^{(m)})} \biggl(\sum_{x^n: \varphi_n(x^n) = u} P_{X^n}(x^n) - \frac{\mathbb{P}[X^n \in \mathcal{D}_m]}{K^m} \biggr) \\
&\!=\! \sum_{u \in \varphi_n(A_n^{(m)})} \sum_{x^n: \varphi_n(x^n) = u} P_{X^n}(x^n) \!-\! \frac{|\varphi_n(A_n^{(m)})| \mathbb{P}[X^n \!\in\! \mathcal{D}_m]}{K^m}  \\
&\ge \sum_{u \in \varphi_n(A_n^{(m)})} \sum_{x^n: \varphi_n(x^n) = u} P_{X^n}(x^n) - \frac{|A_n^{(m)}| \mathbb{P}[X^n \in \mathcal{D}_m]}{K^m}  \\
&\ge \sum_{x^n \in A_n^{(m)}} P_{X^n}(x^n) - \frac{|A_n^{(m)}| \mathbb{P}[X^n \in \mathcal{D}_m]}{K^m}  \\
&=  \sum_{x^n \in A_n^{(m)}} \biggl( P_{X^n}(x^n) - \frac{\mathbb{P}[X^n \in \mathcal{D}_m]}{K^m} \biggr),
\end{align*}
where (k) follows from $V_m \supset \varphi_n(A_n^{(m)})$. Hence, from (\ref{yui1}) and (\ref{yui2}), it follows that
\begin{align}
\sum_{m \in \mathcal{J}(\varphi_n)} \sum_{x^n \in A_n^{(m)}} \biggl( P_{X^n}(x^n) - \frac{\mathbb{P}[X^n \!\in\! \mathcal{D}_m]}{K^m} \biggr) \le \epsilon + \tau. \label{yui3}
\end{align}
From (\ref{yui3}), the definition of $A_n^{(m)}$, and the definition of $\mathcal{Q}_{\epsilon + \tau}(\mathcal{X}^n)$, there exists a $\hat{Q}_{X^n} \in \mathcal{Q}_{\epsilon+\tau}(\mathcal{X}^n)$ such that 
\begin{align}
\hat{Q}_{X^n}(x^n) \le \frac{\mathbb{P}[X^n \in \mathcal{D}_m]}{K^m}, \label{conv3}
\end{align}
for all $m \in \mathcal{J}(\varphi_n)$, and $x^n \in \mathcal{X}^n$ satisfying $\varphi_n(x^n) \in \mathcal{U}^m$. Hence, for $n \ge n_0$, it holds that
\begin{align*}
&\frac{1}{n} G^{[\epsilon + \tau]}(X^n) \\
& \ \overset{(\rm{l})}{\ge} \frac{1}{n} \mathbb{E}_{P_{X^n}} [\iota_{\hat{Q}_{X^n}}(X^n)] \\
& \ = \frac{1}{n} \sum_{m \in \mathcal{J}(\varphi_n)} \sum_{x^n \in \mathcal{D}_m} P_{X^n}(x^n) \iota_{\hat{Q}_{X^n}}(x^n) \\
& \ \overset{(\rm{m})}{\ge} \frac{1}{n} \sum_{m \in \mathcal{J}(\varphi_n)} \sum_{x^n \in \mathcal{D}_m} P_{X^n}(x^n) \log \frac{K^m}{\mathbb{P}[X^n \in \mathcal{D}_m]} \\
& \ \ge \frac{1}{n} \sum_{m \in \mathcal{J}(\varphi_n)} \sum_{x^n \in \mathcal{D}_m} m P_{X^n}(x^n) \\
& \ = \frac{1}{n} \mathbb{E}_{P_{X^n}} [l(\varphi_n(X^n))],
\end{align*}
where (l) follows from $\hat{Q}_{X^n} \in \mathcal{Q}_{\epsilon+\tau}(\mathcal{X}^n)$, (m) follows from (\ref{conv3}). Therefore, it follows that
\begin{align*}
\liminf_{n \to \infty} \frac{1}{n} G^{[\epsilon + \tau]}(X^n) \ge \liminf_{n \to \infty} \frac{1}{n} \mathbb{E}_{P_{X^n}} [l(\varphi_n(X^n))].
\end{align*}
Since $\tau > 0$ is arbitrary, this indicates that
\begin{align}
G^{[\epsilon]}(\mathbf{X}) \ge \liminf_{n \to \infty} \frac{1}{n} \mathbb{E}_{P_{X^n}} [l(\varphi_n(X^n))]. \label{conv5}
\end{align}
By (\ref{hyoukakijun2}) and (\ref{conv5}), $G^{[\epsilon]}(\mathbf{X}) \ge R$. Hence, we have $S_{\rm i}(\epsilon | \mathbf{X}) \le G^{[\epsilon]}(\mathbf{X})$.

\subsection{Proof of Theorem 4}

For any $\gamma>0$, we define $R_0 = \underline{H}_{\epsilon}(\mathbf{X}) - \gamma$. From the definition of $\underline{H}_{\epsilon}(\mathbf{X})$, it holds that
\begin{align}
\limsup_{n \to \infty} \mathbb{P} \biggl[ \frac{1}{n} \iota_{P_{X^n}} (X^n) \le R_0 \biggr] \le \epsilon. \label{tyounaosi1}
\end{align}
For $\tau > 0$, there exists an $n_0 \in \mathbb{N}$ such that
\begin{align}
\mathbb{P} \biggl[ \frac{1}{n} \iota_{P_{X^n}} (X^n) \le R_0 \biggr] \le \limsup_{n \to \infty} \mathbb{P} \biggl[ \frac{1}{n} \iota_{P_{X^n}} (X^n) \le R_0 \biggr] + \tau, \label{tyounaosi2}
\end{align}
for all $n \ge n_0$. From (\ref{tyounaosi1}) and (\ref{tyounaosi2}), for all $n \ge n_0$, it follows that
\begin{align}
\mathbb{P} \biggl[ \frac{1}{n} \iota_{P_{X^n}} (X^n) \le R_0 \biggr] \le \epsilon + \tau. \label{tyounaosi3}
\end{align}

The set $A_n \subset \mathcal{X}^n$ is defined by 
\begin{align*}
A_n = \biggl\{ x^n \in \mathcal{X}^n \mid \frac{1}{n} \iota_{P_{X^n}} (X^n) \le R_0 \biggr\}.
\end{align*}
For $x^n \in A_n$, it holds that
\begin{align}
P_{X^n}(x^n) \ge K^{-nR_0}. \label{tyounaosi4}
\end{align}
Next, we define the sub-probability $\tilde{Q}_{X^n}$ by
\begin{align}
\tilde{Q}_{X^n}(x^n) = 
\begin{cases}
\frac{K^{-nR_0}}{|A_n|}, \ \ \ (x^n \in A_n), \\
P_{X^n}(x^n), \ \ \ (x^n \notin A_n). \label{e-enu}
\end{cases}
\end{align}
From (\ref{tyounaosi4}) and (\ref{e-enu}), for $x^n \in \mathcal{X}^n$, it follows that
\begin{align}
\tilde{Q}_{X^n}(x^n) \le P_{X^n}(x^n), \label{tyounaosi5}
\end{align}
which implies that 
\begin{align}
\frac{1}{n} \iota_{\tilde{Q}_{X^n}}(x^n) = \frac{1}{n} \log \frac{|A_n|}{K^{-nR_0}} = R_0 + \frac{1}{n} \log|A_n|, \label{tyounaosi7}
\end{align}
for $x^n \in A_n$.
On the other hand, for $x^n \notin A_n$, we have
\begin{align}
\frac{1}{n} \iota_{\tilde{Q}_{X^n}}(x^n) = \frac{1}{n} \iota_{P_{X^n}}(x^n) > R_0. \label{tyounaosi8}
\end{align}
From (\ref{tyounaosi7}) and (\ref{tyounaosi8}), for $x^n \in \mathcal{X}^n$, it follows that
\begin{align}
\frac{1}{n} \iota_{\tilde{Q}_{X^n}}(x^n) \ge R_0. \label{tyounaosi9}
\end{align}
Further, for $n \ge n_0$, it holds that
\begin{align}
\sum_{x^n \in \mathcal{X}^n} \tilde{Q}_{X^n}(x^n) 
&= \sum_{x^n \in A_n} \tilde{Q}_{X^n}(x^n) + \sum_{x^n \notin A_n} \tilde{Q}_{X^n}(x^n) \notag \\
&= \sum_{x^n \in A_n} \frac{K^{-nR_0}}{|A_n|} + \sum_{x^n \notin A_n} P_{X^n}(x^n) \notag \\
&= K^{-nR_0} + \mathbb{P} \biggl[ \frac{1}{n} \iota_{P_{X^n}} (X^n) > R_0 \biggr] \notag \\
&\overset{(\rm{n})}{\ge} K^{-nR_0} + 1 - \epsilon - \tau \ge 1 - \epsilon - \tau, \label{tyounaosi10}
\end{align}
where (n) follows from (\ref{tyounaosi3}). From the definition of $\mathcal{Q}_{\epsilon+\tau}(\mathcal{X}^n)$, (\ref{tyounaosi5}), and (\ref{tyounaosi10}), for $n \ge n_0$, there exists a $\bar{Q}_{X^n} \in \mathcal{Q}_{\epsilon + \tau}(\mathcal{X}^n)$ such that
\begin{align}
\iota_{\bar{Q}_{X^n}}(x^n) \ge \iota_{\tilde{Q}_{X^n}}(x^n) \ge \iota_{P_{X^n}}(x^n), \label{tyounaosi6}
\end{align}
for all $x^n \in \mathcal{X}^n$. From (\ref{tyounaosi9}) and (\ref{tyounaosi6}), for any $n \ge n_0$ and $x^n \in \mathcal{X}^n$, it follows that
\begin{align}
\frac{1}{n} \iota_{\bar{Q}_{X^n}}(x^n) \ge R_0. \label{tyounaosi11}
\end{align}

Hence, we have
\begin{align}
\underline{H}_{\epsilon}(\mathbf{X}) - \gamma &= R_0 \overset{(\rm{o})}{\le} \frac{1}{n} \min_{x^n \in \mathcal{X}^n} \iota_{\bar{Q}_{X^n}}(x^n) \notag \\
&= \frac{1}{n} \sum_{x^n \in \mathcal{X}^n} P_{X^n}(x^n) \min_{x^n \in \mathcal{X}^n} \iota_{\bar{Q}_{X^n}}(x^n) \notag \\
&\le \frac{1}{n} \sum_{x^n \in \mathcal{X}^n} P_{X^n}(x^n) \iota_{\bar{Q}_{X^n}}(x^n) \notag \\
&\overset{(\rm{p})}{\le} \frac{1}{n} \sup_{Q_{X^n} \in \mathcal{Q}_{\epsilon+\tau}(\mathcal{X}^n)} \mathbb{E}_{P_{X^n}} [\iota_{Q_{X^n}}(X^n)] \notag \\
&= \frac{1}{n} G^{[\epsilon + \tau]}(X^n), \notag
\end{align}
where (o) follows from (\ref{tyounaosi11}), (p) follows from $\bar{Q}_{X^n} \in \mathcal{Q}_{\epsilon + \tau}(\mathcal{X}^n)$.
Since this formula holds for $n \ge n_0$ and arbitrary $\gamma > 0$ and $\tau > 0$, we have $\underline{H}_{\epsilon}(\mathbf{X}) \le G^{[\epsilon]}(\mathbf{X}).$

\section{Discussion}

In this section, we discuss a duality between the general formula in Theorem 2 and our general formula in Theorem 3 from the viewpoint of the smooth R\'enyi entropy.

The study \cite{koga1} clarified the sub-probability distribution $\mathbf{q}^*$, which achieves the infimum of the {\it smooth R\'enyi entropy of order} $\alpha \in (0,1)$ \cite{renner}. 
In view of the condition of the infimum in Theorem 2, this sub-probability distribution $\mathbf{q}^*$ is related to the sub-probability distribution $\frac{P_{X^n}(x^n)}{\mathbb{P}[X^n \in A_n]}$ of $G_{[\epsilon]}(\mathbf{X})$.

On the other hand, the sub-probability distribution $Q_{X^n} \in \mathcal{Q}_{\epsilon + \tau}(\mathcal{X}^n)$ of $G^{[\epsilon]}(\mathbf{X})$ in Theorem 3 is related to $\mathbf{q}^\dagger$ \cite{koga1}, where $\mathbf{q}^\dagger$ is the sub-probability distribution achieving the infimum of the {\it smooth R\'enyi entropy of order} $\alpha \in (1, \infty)$ \cite{renner}. 

Therefore, we observe a duality between the general formula of $\epsilon$-variable-length resolvability discussed in \cite{yagi} and \cite{yagi2} and that of $\epsilon$-variable-length intrinsic randomness discussed in this paper.

\section{Second-Order Variable-Length Intrinsic Randomness}

We define the {\it problem of $(\epsilon, R)$-variable-length intrinsic randomness}.

\begin{definition}
{\rm Given $\epsilon \in [0,1)$ and $R \ge 0$, a second-order rate $L$ is said to be }{\it i$(\epsilon, R)$-achievable} {\rm if there exists a mapping} $\varphi_n:\mathcal{X}^n \to \mathcal{U}^*$ {\rm satisfying} 
\begin{align*}
\limsup_{n\to \infty} \bar{d}(P_{\varphi_n(X^n)}, P_{U^{(L_n)}}) &\le \epsilon,  \\
\liminf_{n \to \infty} \frac{1}{\sqrt{n}} (\mathbb{E}_{P_{X^n}} [l(\varphi_n(X^n))] - nR) &\ge L. 
\end{align*}
\end{definition}

The $(\epsilon, R)${\it -variable-length intrinsic randomness} is defined as follows.
\begin{definition}
\begin{align*}
T_{\rm i}(\epsilon, R | \mathbf{X}) := \sup \{ L \mid L \ {\rm is } \ {\rm i}(\epsilon, R)\mathchar`-\rm{achievable}\}.
\end{align*}
\end{definition}

We establish the second-order general formula.

\begin{theorem}
For any general source $\mathbf{X}$, 
\begin{align*}
T_{\rm i}(\epsilon, R | \mathbf{X}) = \lim_{\tau \downarrow 0} \liminf_{n \to \infty} \frac{1}{\sqrt{n}} ( &G^{[\epsilon + \tau]}(X^n) -nR ) \\ & \ (\epsilon \in [0, 1), R \ge 0).
\end{align*}
\end{theorem}

\begin{IEEEproof}
The theorem can be proven analogously to Theorem 3 with due modifications. 
\end{IEEEproof}

\section{Conclusion}

We have investigated the problem of $\epsilon$-variable-length intrinsic randomness. The contribution of this paper is to derive the general formula when we allow positive value of the average variational distance and the lower bound of the value characterizing $\epsilon$-variable-length intrinsic randomness. Further, by comparing the previous result by Yagi and Han \cite{yagi}, \cite{yagi2} and our result, we have clarified the dual relationship between the $\epsilon$-variable-length resolvability and the $\epsilon$-variable-length intrinsic randomness.



\appendices

\section{Proof of Lemma 1}

We will use some notations for this proof. 
\begin{itemize}
\item $U_n := U^{(\lfloor naR \rfloor)}$
\item $M_n:= K^{\lfloor naR \rfloor}$
\item $u_i := u_i^{(\lfloor naR \rfloor)}$ \ \ \ $(i = 1, 2, \dots , M_n)$
\end{itemize}
We construct sets $A(i) \subset A_n \ (i=1,2, \dots, M_n)$. First, for $u_1 \in \mathcal{U}^{\lfloor naR \rfloor}$, construct a subset $A(1) \subset A_n$ so as to satisfy the following conditions
\begin{align*}
\sum_{x^n \in A(1)} P_{X^n}(x^n) \le c \cdot P_{U_n}(u_1) = \frac{c}{M_n}
\end{align*}
and, for any $\hat{x}^n \in A_n \setminus A(1)$,
\begin{align*}
c \cdot P_{U_n}(u_1)< \sum_{x^n \in A(1)} P_{X^n}(x^n) + P_{X^n}(\hat{x}^n).
\end{align*}
Next, for $u_2 \in \mathcal{U}^{\lfloor naR \rfloor}$, construct a subset $A(2) \subset A_n \setminus A(1)$ so as to satisfy the following conditions
\begin{align*}
\sum_{x^n \in A(2)} P_{X^n}(x^n) \le c \cdot P_{U_n}(u_2) = \frac{c}{M_n}
\end{align*}
and, for any $\hat{x}^n \in A_n \setminus A(1) \cup A(2)$,
\begin{align*}
c \cdot P_{U_n}(u_2)< \sum_{x^n \in A(2)} P_{X^n}(x^n) + P_{X^n}(\hat{x}^n).
\end{align*}
In an analogous manner, also for $u_3 \in \mathcal{U}^{\lfloor naR \rfloor}$, construct a subset $A(3) \subset A_n \setminus A(1) \cup A(2)$, and so on. Then, $i_0$ is defined as the number of final step of this precedure. For $i_0$, we consider two cases.

\noindent
1) \ case of $i_0 = M_n - 1$:

The set $A(i)\subset A_n \setminus \bigcup_{j=1}^{i-1}A(j) \ (i = 1 , 2, \dots, M_n - 1)$ satisfies the following conditions:
\begin{align}
\sum_{x^n \in A(i)} P_{X^n}(x^n) \le c \cdot P_{U_n}(u_i) = \frac{c}{M_n} \label{natu3}
\end{align}
and, for any $\hat{x}^n \in A_n \setminus \bigcup_{j=1}^i A(j)$,
\begin{align}
c \cdot P_{U_n}(u_i) < \sum_{x^n \in A(i)} P_{X^n}(x^n) + P_{X^n}(\hat{x}^n). \label{2.4}
\end{align}
On the other hand, the set $A(M_n)$ is defined as follows: 
\begin{align*}
A(M_n) := A_n \setminus \bigcup_{i=1}^{M_n - 1}A(i). 
\end{align*} 
From (\ref{2.1}) and (\ref{2.4}), for any $i = 1,2, \dots , M_n - 1$ and $\hat{x}^n \in A_n \setminus \bigcup_{j=1}^i A(j)$, 
\begin{align*}
c \cdot P_{U_n}(u_i) &< \sum_{x^n \in A(i)} P_{X^n}(x^n) + cK^{-n(a+\gamma)R} \\
&= \mathbb{P}[X^n \in A(i)] + cK^{-n(a+\gamma)R}.
\end{align*}
Hence, it holds that
\begin{align}
\mathbb{P}[X^n \in A(i)] > \frac{c}{M_n} - cK^{-n(a+\gamma)R}. \label{natu1}
\end{align}

We define the mapping $\varphi_n: A_n \to \mathcal{U}^{\lfloor naR \rfloor}$ by
\begin{align}
\varphi_n(x^n) = u_i \ \ \ (x^n \in A(i) \ (i=1,2 ,\dots, M_n)). \label{syazou}
\end{align}
Then, it follows that
\begin{align}
&\frac{1}{2} \sum_{u \in \mathcal{U}^{\lfloor naR \rfloor}} \biggl| P_{\varphi_n(X^n)}(u) - \frac{c}{K^{\lfloor naR \rfloor}} \biggr| \notag \\
& \ \le \frac{1}{2} \sum_{i=1}^{M_n} |P_{\varphi_n(X^n)}(u_i) - c \cdot P_{U_n}(u_i)| \notag \\
& \ = \frac{1}{2} \sum_{i=1}^{M_n-1} |P_{\varphi_n(X^n)}(u_i) - c \cdot P_{U_n}(u_i)| \notag \\
& \ \ \ + \frac{1}{2} |P_{\varphi_n(X^n)}(u_{M_n}) -c \cdot P_{U_n}(u_{M_n})|. \label{h1}
\end{align}
We evaluate the second term on the right-hand side of (\ref{h1}).
\begin{align}
&|P_{\varphi_n(X^n)}(u_{M_n}) - c \cdot P_{U_n}(u_{M_n})| \notag \\
&\!=\! \biggl|\mathbb{P}[X^n \!\in\! A_n] - \sum_{i=1}^{M_n - 1} P_{\varphi_n(X^n)}(u_i) \!-\! c \biggl(1\!-\! \sum_{i=1}^{M_n - 1} P_{U_n}(u_i)\biggr)\biggr| \notag \\
&\le (c- \mathbb{P}[X^n \!\in\! A_n]) + \biggl|\sum_{i=1}^{M_n - 1} \biggl(P_{\varphi_n(X^n)}(u_i) - c \cdot P_{U_n}(u_i)\biggr)\biggr| \notag \\
&\!\le\! (c\!-\! \mathbb{P}[X^n \!\in\! A_n]) \!+\! \sum_{i=1}^{M_n - 1} |P_{\varphi_n(X^n)}(u_i) \!-\! c \!\cdot\! P_{U_n}(u_i)|. \label{komachi}
\end{align}
The combination of (\ref{h1}) and (\ref{komachi}) yields
\begin{align}
&\frac{1}{2} \sum_{u \in \mathcal{U}^{\lfloor naR \rfloor}} \biggl| P_{\varphi_n(X^n)}(u) - \frac{c}{K^{\lfloor naR \rfloor}} \biggr| \notag \\
&\le \sum_{i=1}^{M_n-1} |P_{\varphi_n(X^n)}(u_i) - c \!\cdot\! P_{U_n}(u_i)| + \frac{1}{2} (c- \mathbb{P}[X^n \!\in\! A_n]). \label{tuuika0}
\end{align}
Further, we evaluate the first term on the right-hand side of (\ref{tuuika0}).
\begin{align}
&\sum_{i=1}^{M_n-1} |P_{\varphi_n(X^n)}(u_i) - c \cdot P_{U_n}(u_i)| \notag \\
& \ \overset{(\rm{q})}{=} \sum_{i=1}^{M_n-1} |\mathbb{P}[X^n \in A(i)] - c \cdot P_{U_n}(u_i)|  \notag \\
& \ \overset{(\rm{r})}{=} \sum_{i=1}^{M_n-1} \biggr(\frac{c}{M_n} - \mathbb{P}[X^n \in A(i)] \biggr)  \overset{(\rm{s})}{<} \sum_{i=1}^{M_n-1} cK^{-n(a+\gamma)R} \notag \\
& \ \!\le\! M_n cK^{-n(a+\gamma)R}  \!=\! K^{\lfloor naR \rfloor} cK^{-n(a+\gamma)R} \!\le\! cK^{-n\gamma R}, \label{tuuika1}
\end{align} 
where (q) follows from (\ref{syazou}), (r) follows from (\ref{natu3}), (s) follows from (\ref{natu1}). By substituting (\ref{tuuika1}) for (\ref{tuuika0}), we have
\begin{align}
&\frac{1}{2} \sum_{u \in \mathcal{U}^{\lfloor naR \rfloor}} \biggl| P_{\varphi_n(X^n)}(u) - \frac{c}{K^{\lfloor naR \rfloor}} \biggr| \notag \\
& \ \le cK^{-n\gamma R} + \frac{1}{2} (c- \mathbb{P}[X^n \in A_n]). \label{kansei1}
\end{align}

\noindent
2) \ case of $i_0 < M_n - 1$:

The set $A(i)\subset A_n \setminus \bigcup_{j=1}^{i-1}A(j) \ (i = 1 , 2, \dots, i_0)$ satisfies the following conditions:
\begin{align*}
\sum_{x^n \in A(i)} P_{X^n}(x^n) \le c \cdot P_{U_n}(u_i) = \frac{c}{M_n} 
\end{align*}
and, for any $\hat{x}^n \in A_n \setminus \bigcup_{j=1}^i A(j)$, 
\begin{align*}
c \cdot P_{U_n}(u_i) < \sum_{x^n \in A(i)} P_{X^n}(x^n) + P_{X^n}(\hat{x}^n). 
\end{align*}
On the other hand, the set $A(i_0+1)$ is defined as follows:
\begin{align*}
A(i_0+1) := A_n \setminus \bigcup_{i=1}^{i_0} A(i).
\end{align*}
Moreover, define the set $A(i) \ (i = i_0 + 2, \dots ,M_n)$ by
\begin{align}
A(i) := \phi. \label{te-se-6}
\end{align}
Then, for any $i = 1,2, \dots , M_n$, it holds that
\begin{align}
\mathbb{P}[X^n \in A(i)] \le c \cdot P_{U_n}(u_i) = \frac{c}{M_n}. \label{te-se-7}
\end{align}

We define the mapping $\varphi_n: A_n \to \mathcal{U}^{\lfloor naR \rfloor}$ by
\begin{align}
\varphi_n(x^n) = u_i \ \ \ (x^n \in A(i) \ (i=1,2 ,\dots, M_n)). \label{te-se-4}
\end{align}
Then, it follows that
\begin{align}
&\frac{1}{2} \sum_{u \in \mathcal{U}^{\lfloor naR \rfloor}} \biggl| P_{\varphi_n(X^n)}(u) - \frac{c}{K^{\lfloor naR \rfloor}} \biggr| \notag \\
& \ \le \frac{1}{2} \sum_{i=1}^{M_n-1} |P_{\varphi_n(X^n)}(u_i) - c \cdot P_{U_n}(u_i)| \notag \\
& \ \ \ + \frac{1}{2} |P_{\varphi_n(X^n)}(u_{M_n}) -c \cdot P_{U_n}(u_{M_n})|. \label{te-se-3}
\end{align}
We evaluate the first term on the right-hand side of (\ref{te-se-3}).
\begin{align}
&\frac{1}{2} \sum_{i=1}^{M_n-1} |P_{\varphi_n(X^n)}(u_i) - c \cdot P_{U_n}(u_i)| \notag \\ & \ \overset{(\rm{t})}{=} \frac{1}{2} \sum_{i=1}^{M_n-1} |\mathbb{P}[X^n \in A(i)] - c \cdot P_{U_n}(u_i)| \notag \\
& \ \overset{(\rm{u})}{=} \frac{1}{2} \sum_{i=1}^{M_n-1} (c \cdot P_{U_n}(u_i) - \mathbb{P}[X^n \in A(i)]) \notag \\
& \ \overset{(\rm{v})}{=} \frac{c(M_n - 1)}{2M_n} - \frac{1}{2}\mathbb{P}[X^n \in A_n], \label{te-se-10}
\end{align}
where (t) follows from (\ref{te-se-4}), (u) follows from (\ref{te-se-7}), (v) follows from the construction of $A(i)$.
Next, we evaluate the second term on the right-hand side of (\ref{te-se-3}). From the analogous calculation of the first term on the right-hand side of (\ref{te-se-3}), it holds that
\begin{align}
&\frac{1}{2} |P_{\varphi_n(X^n)}(u_{M_n}) -c \cdot P_{U_n}(u_{M_n})| \notag \\
& \ = \frac{1}{2} |\mathbb{P}[X^n \in A(M_n)] - c \cdot P_{U_n}(u_{M_n})| \notag \\
& \ = \frac{1}{2} (c \cdot P_{U_n}(u_{M_n}) - \mathbb{P}[X^n \in A(M_n)]) \overset{(\rm{w})}{=} \frac{c}{2M_n}, \label{te-se-9}
\end{align}
where (w) follows from $i_0 + 1 < M_n$ and (\ref{te-se-6}). By substituting (\ref{te-se-10}) and (\ref{te-se-9}) for (\ref{te-se-3}), we have
\begin{align}
&\frac{1}{2} \sum_{u \in \mathcal{U}^{\lfloor naR \rfloor}} \biggl| P_{\varphi_n(X^n)}(u) - \frac{c}{K^{\lfloor naR \rfloor}} \biggr| \notag \\
&\le \frac{c(M_n \!-\! 1)}{2M_n} - \frac{1}{2}\mathbb{P}[X^n \!\in\! A_n]  +  \frac{c}{2M_n} = \frac{1}{2} (c\!-\! \mathbb{P}[X^n \!\in\! A_n]). \label{kansei2}
\end{align}
By (\ref{kansei1}) and (\ref{kansei2}), we can prove (\ref{hodai1}).

\section*{Acknowledgment}

The authors would like to thank Dr.\ Hideki Yagi and Dr.\ Ryo Nomura for helpful discussions.
This work was supported in part by JSPS KAKENHI Grant Numbers JP16K00195, JP16K00417, JP17K00316, JP17K06446, and by Waseda University Grant for Special Research Projects (Project number: 2017A-022).



%

\end{document}